\documentclass[aps,preprint,superscriptaddress,showpacs]{revtex4}
\usepackage{graphicx}
\usepackage{dcolumn}
\usepackage{amsmath}
\usepackage{tabularx}

\begin{document}

\bibliographystyle{apsrev}

\preprint{BARI-TH 407/2001}

\title{Lattice measurement of the energy-gap \\
in a spontaneously broken phase}

\author{P. Cea}
\email[]{Paolo.Cea@ba.infn.it}
\affiliation{INFN - Sezione di Bari, via Amendola 173, 
I 70126 Bari, Italy}
\affiliation{Dipartimento di Fisica,  Universit\`a di Bari,  via Amendola 173, 
I 70126 Bari, Italy}
\author{M. Consoli}
\email[]{Maurizio.Consoli@ct.infn.it}
\affiliation{INFN - Sezione di Catania, Corso Italia 57, I 95129 Catania, 
Italy}
\author{L. Cosmai}
\email[]{Leonardo.Cosmai@ba.infn.it}
\affiliation{INFN - Sezione di Bari, via Amendola 173, 
I 70126 Bari, Italy}

\date{March 21, 2001}

\begin{abstract}
Using lattice simulations of a one-component
$(\lambda \Phi^4)_4$ theory, we have measured the energy spectrum
$\omega({\mathbf{k}})$ 
in the broken phase at various lattice sizes.
Our data show that the energy-gap 
$\omega(0)$ is {\it not} the `Higgs mass' $M_h$ but an infrared-sensitive
quantity that becomes smaller and smaller by increasing the lattice 
size and may even vanish in the infinite-volume limit. 
\end{abstract}
\maketitle

In the case of a one-component $\lambda \Phi^4$ theory, 
and due to the underlying `triviality' of the theory in 3+1 space-time
dimensions \cite{book}, the energy spectrum of the broken symmetry phase 
$\omega( {\mathbf{k}}) $ is believed to 
approach a single-particle form, say
$\sqrt{  {\mathbf{k}}^2 + M^2_h} $, in the continuum limit of quantum field theory.
Equivalently, the energy-gap $\omega(0)$ is assumed to represent a good measure
of the `Higgs mass' $M_h$. This statement
is explicitely supported by the analysis of 
Ref.~\cite{Luscher:1988ek}, where all perturbative ambiguities
in the definition of the Higgs mass
are shown to be very small in the scaling region.

Thus, by approaching the continuum limit with
a lattice simulation of the theory, 
the shape of the energy spectrum should be better and better 
reproduced by (the lattice version of) a single-particle form
$\sqrt{  {\mathbf{k}}^2 + {\text{const}}} $. 
Of course, this is completely equivalent to 
a continuum limit of the shifted-field propagator 
\begin{equation}
\label{gpform}
G(p) \to \frac{Z_{\text{ prop}}}{ {p}^2 + \bar{m}^2 } ,
\end{equation}
with $Z_{\text{ prop}} \to 1$ for consistency with the K\"allen-Lehmann 
spectral decomposition of a `trivial' theory. 

These theoretical expectations have been compared with the results of 
model-independent
lattice simulations in Refs.~\cite{Cea:1998hy,Cea:1999kn}. The lattice data 
for the scalar propagator are well reproduced by 
Eq.~(\ref{gpform}) in the
symmetric phase. In the broken phase, however, the fit 
with Eq.~(\ref{gpform}), although excellent at high momentum, becomes very
poor for $p \to 0$. Equivalently, 
the measured energy spectrum is well reproduced by the single-particle form
$\sqrt{  {\mathbf{k}}^2 +  \bar{m}^2 } $ for not too small $|{\mathbf{k}}|$.
However, the spectrum 
deviates significantly when ${\mathbf{k}} \to 0$ and
direct measurements of
 $\omega(0)$ exhibit larger and larger percentage deviations from
$M_h \equiv \bar{m}$ by approaching the continuum limit 
\cite{Cea:1999kn}.
Motivated by these unexpected discrepancies, we have undertaken a more systematic
analysis of the energy spectrum on various $L^4$ lattices with
$20 \leq L \leq 40$.  Our results will be reported in the following.

The one-component $(\lambda\Phi^4)_4$ theory   
\begin{equation}
\label{action}
   S =\sum_x \left[ \frac{1}{2}\sum_{\mu}(\Phi(x+\hat e_{\mu}) - 
\Phi(x))^2 + \frac{r_0}{2}\Phi^2(x)  + \frac{\lambda_0}{4} \Phi^4(x)  
\right]    
\end{equation}
is conveniently studied in the Ising limit 
\begin{equation}
\label{ising}
   S_{\text{ Ising}} = -\kappa
\sum_x\sum_{\mu} \left[ 
\phi(x+\hat e_{\mu})\phi(x) +
\phi(x-\hat e_{\mu})\phi(x) \right]    
\end{equation}
with $\Phi(x)=\sqrt{2\kappa}\phi(x)$ and where $\phi(x)$ takes only the 
values $+1$ or $-1$.  The broken phase is found for values 
$\kappa > \kappa_c \sim 0.0748$ \cite{montmunster}.

    We performed Monte-Carlo simulations of this Ising action 
using the Swendsen-Wang \cite{Swendsen:1987ce} cluster algorithm.  
Statistical errors 
can be estimated through a direct evaluation of the integrated autocorrelation 
time~\cite{Madras:1988ei}, or by using the 
``blocking''~\cite{Whitmer:1984he,Flyvbjerg:1989} or the 
``grouped jackknife''~\cite{Efron:1982,Berg:1989cp} algorithms.  
We have checked that 
applying these three different methods we get consistent results.  

    As an approach to the `Higgs mass' we have used the method of ``time-slice'' 
variables described in Ref.~\cite{Montvay:1987us} 
(see also \cite{montmunster} pp. 56) which has the advantage of being
independent of uncontrolled theoretical assumptions.
To this end, let us consider a lattice with 3-dimension $L^3$ and 
temporal dimension $L_t$ and the two-point correlator 
\begin{equation}
\label{corr}
C_1(t,0; {\mathbf k})\equiv \langle 
S_c(t;{\mathbf k})S_c(0;{\mathbf k})+
S_s(t;{\mathbf k})S_s(0;{\mathbf k}) \rangle _{\text{ conn}} ,
\end{equation}
where
\begin{equation}
\label{cos}
S_c(t; {\mathbf k})\equiv \frac{1}{L^3} \sum _{ { \mathbf x} } \phi({\mathbf x}, t)
\cos ({\mathbf k} \cdot {\mathbf x}) ,
\end{equation}
\begin{equation}
\label{sin}
S_s(t;{ \mathbf k})\equiv \frac{1}{L^3} \sum _ {{\mathbf x}} \phi({\mathbf x}, t)
\sin ({\mathbf k} \cdot {\mathbf x}) .
\end{equation}
Here, $t$ is the Euclidean time; ${\mathbf x}$ is the spatial part of the site 
4-vector $x^{\mu}$; ${\mathbf k}$ is the lattice momentum 
${\mathbf k}=(2\pi/L) (n_x,n_y,n_z$), with $(n_x,n_y,n_z)$ non-negative integers; 
and $\langle ...\rangle_{\text{ conn}}$ denotes the connected expectation 
value with respect to the lattice action, Eq.~(\ref{ising}). In this way, 
parameterizing the correlator $C_1$ in terms of the energy $\omega({\mathbf k})$ as 
\begin{equation}
\label{fitcor}
C_1(t,0;{\mathbf k})= A \, [ \, \exp(-\omega({\mathbf k}) 
 t)+\exp(-\omega({\mathbf k}) (L_t-t)) \, ] \,,
\end{equation}
a mass can be defined through the lattice single-particle dispersion relation
\begin{equation}
\label{disp}
m^2_{\text{ TS}}
({\mathbf k}) 
= ~2 (\cosh \omega({\mathbf k})   -1)~~ -~~2 \sum ^{3} _{\mu=1}~ 
(1-\cos k_\mu) \,.
\end{equation}
In the broken phase, by adopting Eq.~(\ref{fitcor}) one neglects the effect of tunneling 
between the two degenerate vacua. For the value of $\kappa$ that we shall consider
the tunneling effect is negligible~\cite{Jansen:1989cw} for lattices as large as $20^4$,
as in our case (see also the Appendix of Ref.~\cite{Cea:1999kn}).
In a massive free-field theory $m_{\text{ TS}}$
is independent of ${\mathbf k}$ and 
coincides with the mass from Eq.~(\ref{gpform}). In general, observable
deviations of $m_{\text{TS}}$ from a simple constant behaviour give a measure
of those contributions to the energy spectrum that go beyond a
single-particle form. However, regardless of any theoretical model, 
$m_{\text{TS}}(0)$ defines $\omega(0)$, the energy-gap of the theory.

As a check of our simulations we started our analysis at  $\kappa=0.0740$
in the symmetric phase on a $20^4$ lattice, where 
high-statistics results by Montvay and 
Weisz~\cite{Montvay:1987us} are available.   
In Fig.~1 we show the values of the time-slice mass $m_{\text{ TS}}({\mathbf k})$ 
(Eq.~(\ref{disp})) at 
several values of the  3-momentum. The shaded area corresponds to the value 
$ \bar{m} = 0.2141(28)$ obtained from the fit to the propagator data 
in Ref.~\cite{Cea:1999kn} and perfectly agrees with the result of Ref.~\cite{Montvay:1987us}.
We see that $m_{\text{ TS}}$ is indeed independent of ${\mathbf k}$ 
so that the energy spectrum of the symmetric phase is very
well reproduced by a single-particle form as expected.
Finally, we have checked two
values of ${\mathbf{k}}$ on a bigger $32^4$ lattice. Notice that, 
even for a lattice mass as small as $0.2$, the numerical 
value of the energy-gap remains remarkably stable. 

We now choose for $\kappa$ the value
$\kappa = 0.076$ in the broken symmetry phase where
high-statistics results by Jansen et al. 
\cite{Jansen:1989cw} are available. 
In this case, the time-slice mass $ m_{\text{ TS}}({\mathbf k})$ shows 
a distinctive behaviour as seen in Fig.~2. At higher momentum, 
the time-slice mass agrees well with the value of $M_h=\bar{m}$ obtained 
in Ref.~\cite{Cea:1999kn}
from a fit to the propagator data at high momenta. On the other hand, 
there are sizeable deviations when ${\mathbf{k}} \to 0$. 

The most striking result concerns, however, 
the time-slice mass at zero momentum. 
In Table 1 we have reported the outputs
of several independent lattice simulations obtained from different random 
sequences (we used the pseudorandom numbers generator 
RANLUX~\cite{Luscher:1994dy,James:1994vv} with `luxury level' 4).
Regardless of the operative definition adopted for the `Higgs mass', 
the energy-gap itself becomes smaller 
and smaller by increasing the lattice size (see Fig.~3).
This result should be compared with
the remarkable stability of $M_h\equiv \bar{m}$, 
as extracted from the set of the high-momentum
data (see Table 2 of Ref.~\cite{Cea:1999kn}). As one can check, 
Eq.~(\ref{fitcor}) provides an excellent fit to the lattice data (see Fig.~4).

Thus, our lattice simulation shows that the energy-gap {\it in the broken
phase} is an infrared-sensitive quantity that becomes smaller and smaller 
by increasing the lattice size and may even vanish in the infinite-volume limit. 
Quite independently of the Goldstone phenomenon,
this may signal the existence of long-wavelength collective excitations 
of the scalar condensate.
This would explain why $\omega(0)$ {\it cannot} be taken as the input definition
of $M_h$ that, rather, has to be extracted from those values of $\omega(\mathbf{k})$
that are well reproduced by the single-particle 
form $\sqrt{ {\mathbf{k}}^2 + {\text{ const}} }$.

If the energy-gap $\omega(0)$ 
vanishes in an infinite volume, as our data
suggest, the same conclusion holds in a spontaneously broken
continuous O(N) symmetry for the energy spectrum of the singlet
Higgs field.
Therefore in the Standard Model there would
be unexpected long-range forces that survive after coupling the scalar
fields to gauge bosons. In view of the importance of the issue, we hope and
expect that our numerical
results for the energy-gap
will be checked (and/or challenged) by other groups.

\newpage
\begin{table}
\caption{The energy-gap, the magnetization, and the susceptibility for $\kappa=0.076$ at various lattice sizes. 
The reported data refer to independent simulations.}
\label{Table1}
\begin{ruledtabular}
\begin{tabular*}{\hsize}{l@{\extracolsep{0ptplus1fil}}r@{\extracolsep{0ptplus1fil}}
l@{\extracolsep{0ptplus1fil}}l@{\extracolsep{0ptplus1fil}}l}
lattice size &  \#configs. & $\omega(0)$ &  $<|\phi|>$ &   $\chi$  \\
\colrule
$20^4$           &  7500K      & 0.3912(12)\footnotemark[1]   & 0.30158(2)\footnotemark[1]   & 37.85(6)\footnotemark[1]  \\
$24^4$           &  3950K      & 0.3820(47)                   & 0.301592(20)                 & 37.66(8)                  \\
$24^4$           &  2750K      & 0.3756(53)                   & 0.301594(24)                 & 37.55(9)                  \\
$32^4$           &   820K      & 0.3438(125)                  & 0.301567(31)                 & 37.74(21)                 \\
$32^4$           &   620K      & 0.3558(150)                  & 0.301569(28)                 & 37.72(28)                 \\
$32^4$           &  1000K      & 0.3353(135)                  & 0.301593(27)                 & 37.73(20)                 \\
$40^4$           &   375K      & 0.2940(182)                  & 0.301564(24)                 & 37.69(37)                 \\
$40^4$           &   240K      & 0.3051(262)                  & 0.301601(35)                 & 38.13(32)                 \\
\end{tabular*}
\end{ruledtabular}
\footnotetext[1]{from Ref.~\onlinecite{Jansen:1989cw}}
\end{table}

\newpage
\begin{figure}
\includegraphics[clip,width=\hsize]{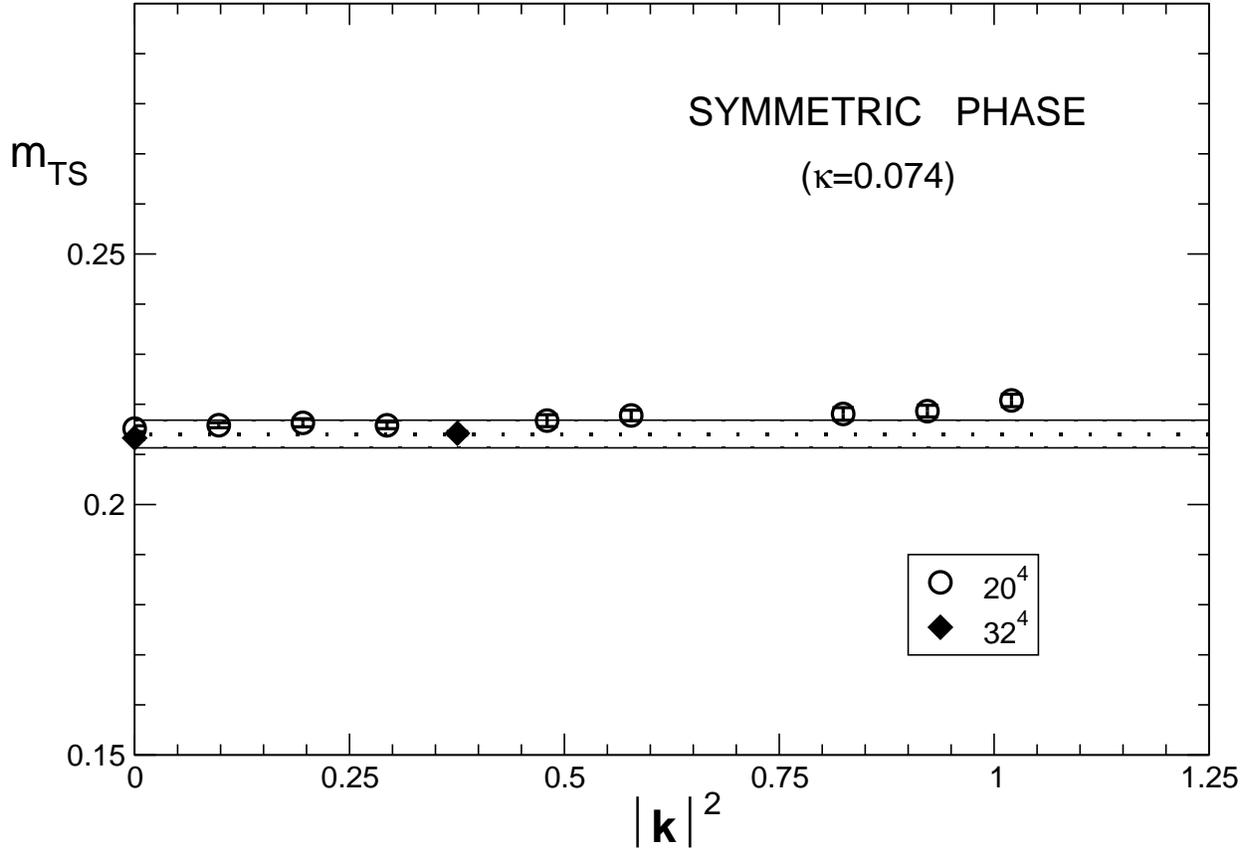}
\caption{The data for the time-slice mass Eq.~(\ref{disp}) 
at different values of the
3-momentum. The shaded area represents the value $\bar{m} = 0.2141(28)$
obtained in Ref.~\cite{Cea:1999kn} from the fit to the propagator data and
perfectly agrees with the value $0.2125(10)$ of Ref.~\cite{Montvay:1987us}.}
\label{fig01}
\end{figure}

\newpage
\begin{figure}
\includegraphics[clip,width=\hsize]{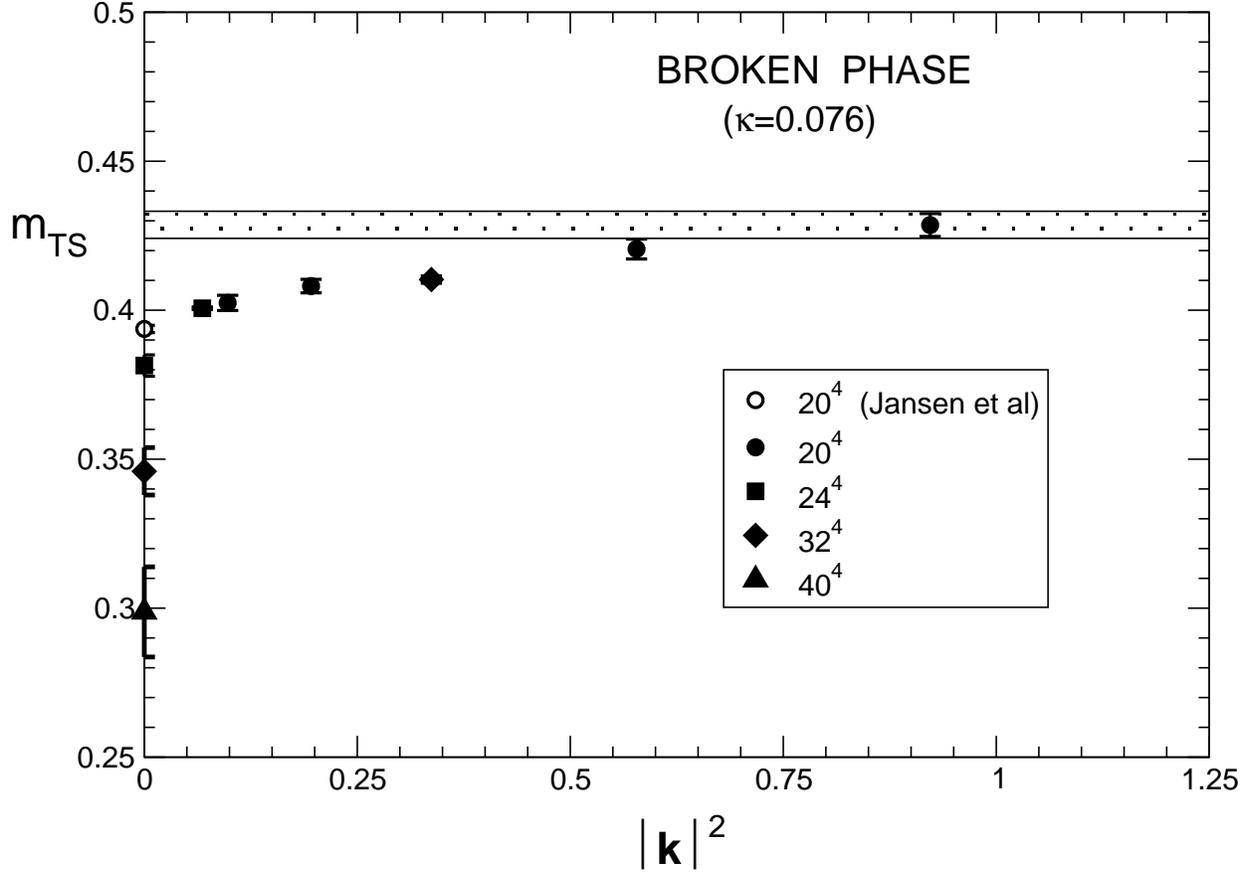}
\caption{The time-slice mass Eq.~(\ref{disp})
for several values of the spatial momentum and different lattice sizes.  
The open circle at zero momentum is the result of Ref.~\cite{Jansen:1989cw}.
Our zero-momentum values are weighted averages of the corresponding measurements
shown in Table~\ref{Table1}.  
The shaded area represents the value 
$\bar{m}=0.42865(456)$ obtained from the propagator data that are well
fitted by Eq.~(\ref{gpform}), see Ref.~\cite{Cea:1999kn}.}
\label{fig02}
\end{figure}

\newpage
\begin{figure}
\includegraphics[clip,width=0.9\hsize]{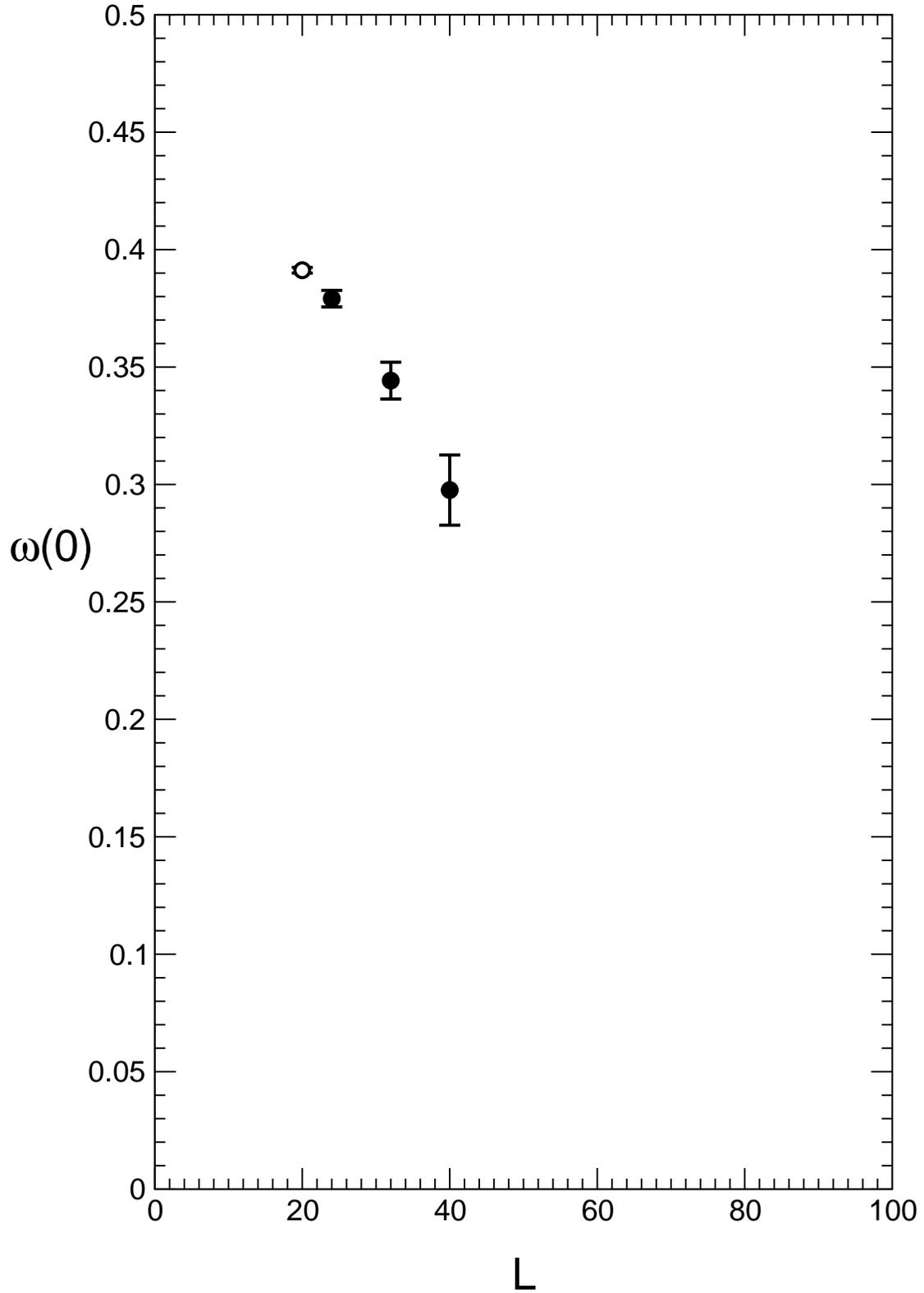}
\caption{The measured energy-gap for $\kappa=0.076$ at different lattice sizes. 
The reported values are
the weighted averages of the results in Table~\ref{Table1}. The value for $L=20$ is
from Ref.~\cite{Jansen:1989cw}.}
\label{fig03}
\end{figure}

\newpage
\begin{figure}
\includegraphics[clip,width=0.9\hsize]{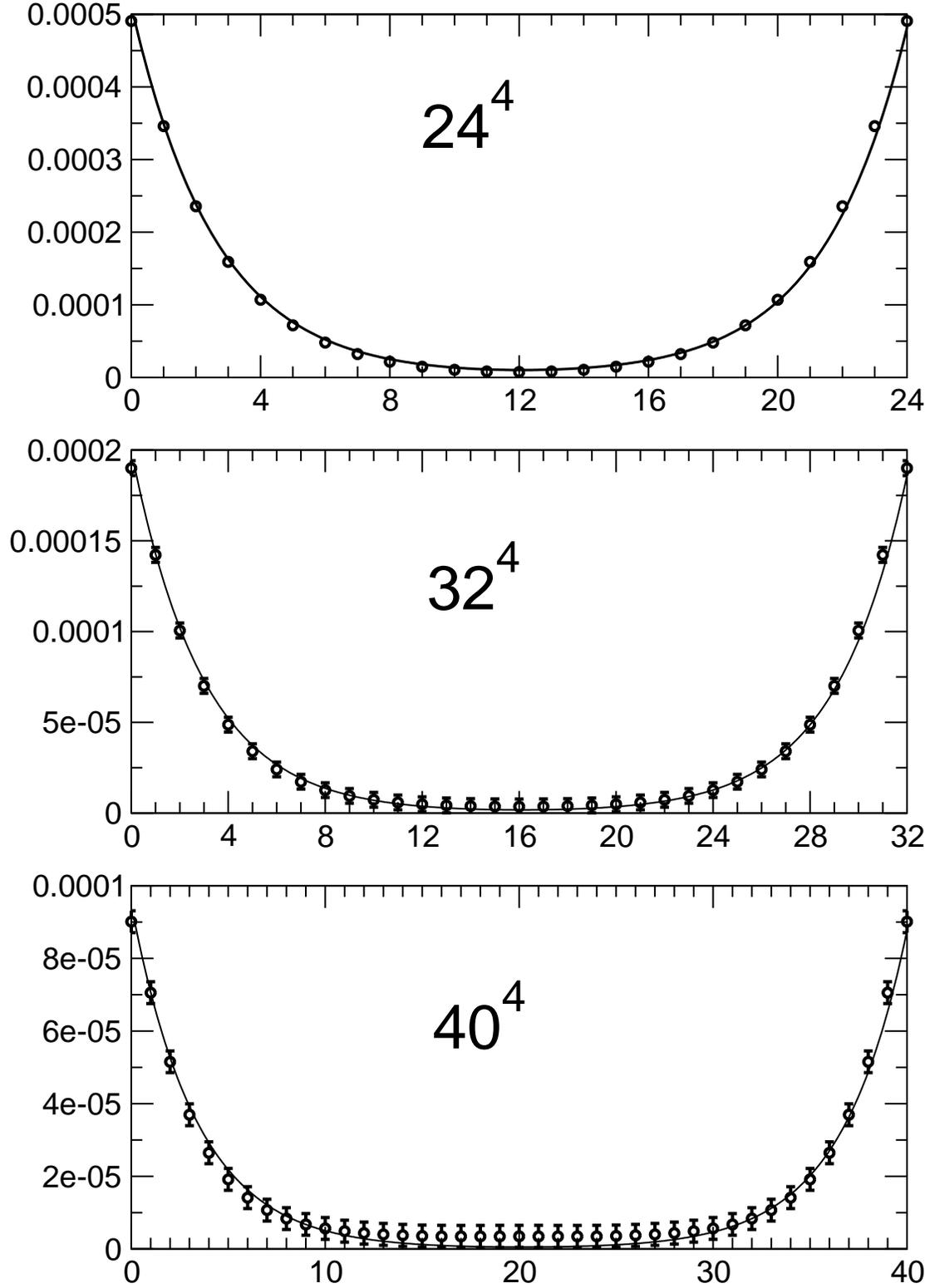}
\caption{The lattice data for the connected correlator Eq.~(\ref{corr}) 
as a function of $t$ at $\mathbf{k}=0$.  
The reported data refer to $24^4$, $32^4$, and $40^4$ lattices in the broken phase at $\kappa=0.076$.
The solid line is the fit with Eq.~(\ref{fitcor}).}
\label{fig04}
\end{figure}


\end{document}